\documentclass[twocolumn,showpacs,superscriptaddress, preprintnumbers , amsmath,amssymb,letterpaper]{revtex4-1}

\usepackage{lineno}

\usepackage{dcolumn}
\usepackage{bm}
\usepackage{amsmath}
\usepackage{multirow}
\usepackage{graphicx}
\usepackage{float}
\usepackage[colorlinks=true]{hyperref}

\begin{document}

 \newcommand{\re}{\mathop{\mathrm{Re}}}
 \newcommand{\im}{\mathop{\mathrm{Im}}}
 \newcommand{\D}{\mathop{\mathrm{d}}}
 \newcommand{\I}{\mathop{\mathrm{i}}}
 \newcommand{\E}{\mathop{\mathrm{e}}}
 \newcommand{\unite}[2]{\mbox{$#1\,{\rm #2}$}}
 \newcommand{\myvec}[1]{\mbox{$\overrightarrow{#1}$}}
 \newcommand{\mynor}[1]{\mbox{$\widehat{#1}$}}
\newcommand{\rmsemit}{\mbox{$\tilde{\varepsilon}$}}
\newcommand{\mean}[1]{\mbox{$\langle{#1}\rangle$}}
\newcommand{\warp}{{\sc Warp }}
\newcommand{\astra}{{\sc astra }}
\newcommand{\elegant}{{\sc Elegant }}
\newcommand{\astragenerator}{{\sc AstraGenerator }}
\newcommand{\mafia}{{\sc Mafia }}

\title{Passive Ballistic Microbunching of Non-Ultrarelativistic Electron Bunches 
using Electromagnetic Wakefields in Dielectric-Lined Waveguides}
\author{F.  Lemery} 
\affiliation{Deutsches Elektronen-Synchrotron, Notkestra\ss e 85, 22607 Hamburg, Germany} 
\author{P.  Piot} 
\affiliation{Northern Illinois Center for Accelerator \& Detector Development and Department of Physics, Northern Illinois University, DeKalb IL 60115, USA} 
\affiliation{Accelerator Physics Center, Fermi National Accelerator Laboratory, Batavia, IL 60510, USA}
\author{G. Amatuni} 
\affiliation{Deutsches Elektronen-Synchrotron, Platannenallee 6, 15738 Zeuthen, Germany} 
\affiliation{Center for the Advancement of Natural Discoveries using Light Emission, Yerevan, Armenia} 
\author{P.  Boonpornprasert}
\affiliation{Deutsches Elektronen-Synchrotron, Platannenallee 6, 15738 Zeuthen, Germany} 
\author{Y. Chen}
\affiliation{Deutsches Elektronen-Synchrotron, Platannenallee 6, 15738 Zeuthen, Germany} 
\author{J. Good}
\affiliation{Deutsches Elektronen-Synchrotron, Platannenallee 6, 15738 Zeuthen, Germany} 
\author{B. Grigoryan}
\affiliation{Deutsches Elektronen-Synchrotron, Platannenallee 6, 15738 Zeuthen, Germany} 
\affiliation{Center for the Advancement of Natural Discoveries using Light Emission, Yerevan, Armenia} 
\author{M. Gro\ss}
\affiliation{Deutsches Elektronen-Synchrotron, Platannenallee 6, 15738 Zeuthen, Germany} 
\author{M. Krasilinikov}
\affiliation{Deutsches Elektronen-Synchrotron, Platannenallee 6, 15738 Zeuthen, Germany} 
\author{O.  Lishilin}
\affiliation{Deutsches Elektronen-Synchrotron, Platannenallee 6, 15738 Zeuthen, Germany} 
\author{G.  Loisch}
\affiliation{Deutsches Elektronen-Synchrotron, Platannenallee 6, 15738 Zeuthen, Germany} 
\author{A. Oppelt}
\affiliation{Deutsches Elektronen-Synchrotron, Platannenallee 6, 15738 Zeuthen, Germany} 
\author{S. Philipp}
\affiliation{Deutsches Elektronen-Synchrotron, Platannenallee 6, 15738 Zeuthen, Germany} 
\author{H. Qian}
\affiliation{Deutsches Elektronen-Synchrotron, Platannenallee 6, 15738 Zeuthen, Germany} 
\author{Y. Renier}
\affiliation{Deutsches Elektronen-Synchrotron, Platannenallee 6, 15738 Zeuthen, Germany} 
\author{F. Stephan} 
\affiliation{Deutsches Elektronen-Synchrotron, Platannenallee 6, 15738 Zeuthen, Germany} 
\author{I. Zagorodnov} 
\affiliation{Deutsches Elektronen-Synchrotron, Notkestra\ss e 85, 22607 Hamburg, Germany} 

%


\date{\today}

\begin{abstract}
Temporally-modulated electron beams have a wide array of applications ranging from the generation of coherently-enhanced electromagnetic radiation to the resonant excitation of electromagnetic wakefields in advanced-accelerator concepts. Likewise producing low-energy ultrashort microbunches could be useful for ultra-fast electron diffraction and new accelerator-based light-source concepts. In this Letter we propose and experimentally demonstrate a passive microbunching technique capable of forming a picosecond bunch train at $\sim 6$~MeV. 
The method relies on the  excitation of electromagnetic wakefields as the beam propagates through a dielectric-lined waveguide. Owing to the non-ultrarelativistic nature of the beam, the induced energy modulation eventually converts into a density modulation as the beam travels in a following free-space drift. The modulated beam is further accelerated to $\sim20$~MeV while preserving the imparted density modulation.  
\end{abstract}
\maketitle
Forefront applications of electron beams call for increasingly precise spatio-temporal control over the beam phase-space distribution. Beam-manipulation techniques to tailor the distributions of electron bunches have flourished over the last decade and include various degrees of complexity~\cite{englandPRL,muggliPRL,sunPRL,piotPRL,haPRL}. 
Recently, methods to passively shape the temporal (or current) distribution of an electron beam have emerged~\cite{antipov,lemery,andonian}. In essence, this class of techniques uses a dielectric-lined waveguide (DLW) to impart an arbitrary time-energy correlation along an electron bunch; subsequently a suitable beamline converts the induced energy correlations into a temporal distribution i.e. current profile. The techniques successfully demonstrated so far~\cite{antipov,andonian} were realized at relativistic energies and use a dispersive section composed of a magnetic chicane~\cite{carlsten} to manipulate the current profile.

In this Letter we demonstrate that a DLW located directly downstream of a photoemission electron source supports the formation of a current-modulated beam over a drift in free space, thereby avoiding a magnetic-based dispersive section and associated dilution of the phase-space distribution in the bending-plane degree of freedom~\cite{BC}. The formed current-modulated beams could be injected in a subsequent linear accelerator to allow for further tailoring. Additionally, the availability of shaped low-energy modulated beams~\cite{piotFEL15} could have direct application to THz light sources~\cite{gover,XFELTHz} or ultra-fast electron diffraction~\cite{ued0,ued}.  

\begin{figure}[hhhh!!!!!!!!!!!!]
\centering
\includegraphics[width=0.47\textwidth]{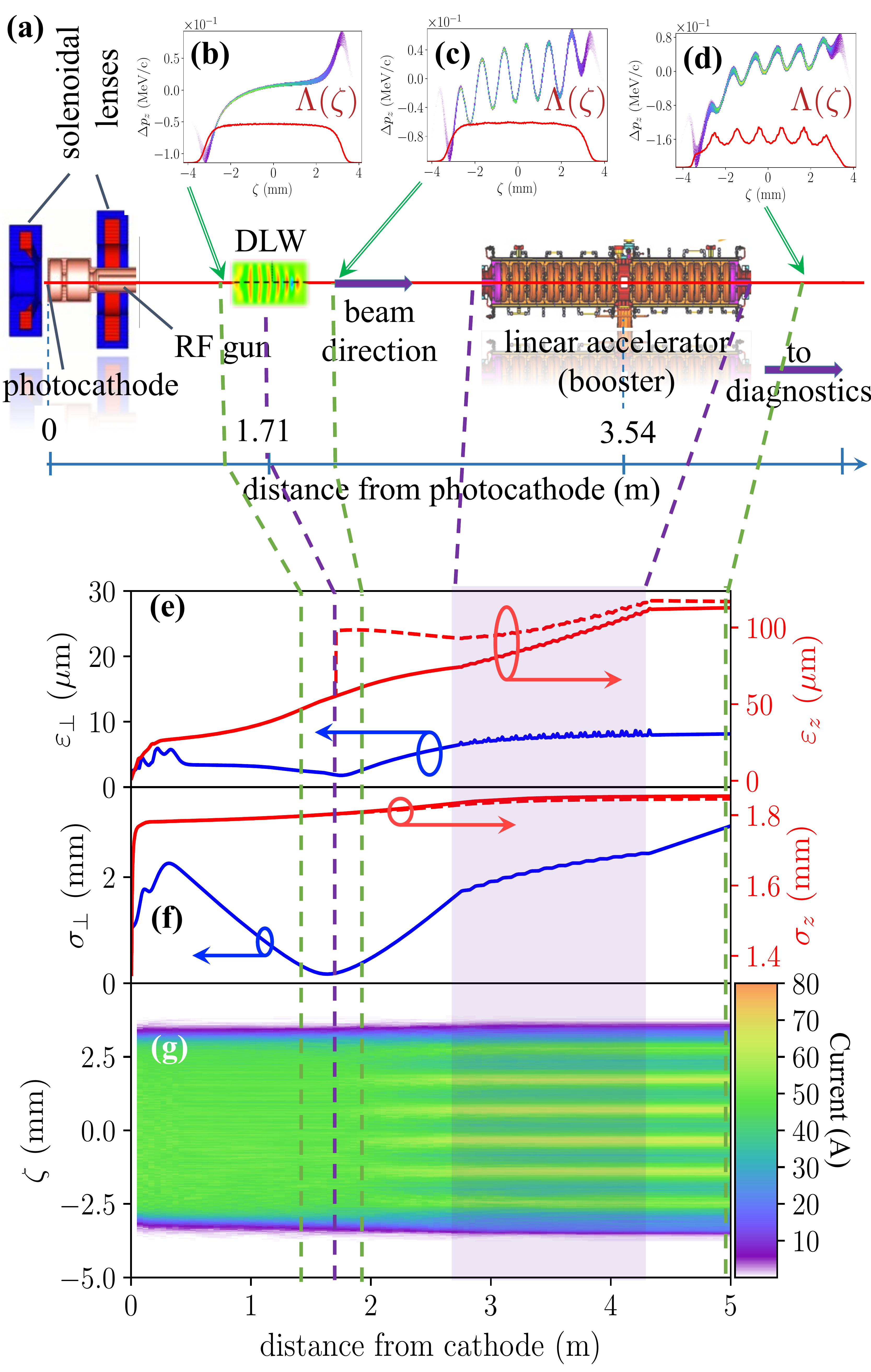}
\caption{Overview of the passive ballistic-bunching experiment implemented at the PITZ facility (a), simulated evolutions of the longitudinal-phase-space density distribution $(\zeta,\delta)$ at the different stages of the bunching process (b-d) with associated current profile [$\Lambda(\zeta)$], evolution of the transverse and longitudinal emittances (e) and rms beam size (f) along the accelerator beamline with (dashed trace) and without (solid trace) DLW2 present, and development of the bunch current profile [$I(\zeta)$] along the beamline (g). Note that in plots (e) and (f) the dashed and solid traces overlap for the transverse parameters. The nominal parameters for these simulations are listed in Tab.~\ref{tab:settings} for the case of DLW2. Values $\zeta>0$ correspond to the head of the bunch. }
\label{fig:setup}
\end{figure}

In order to quantify the proposed self-bunching mechanism, we model the electron bunch as a line-charge distribution and analyze the dynamics of the electrons in the longitudinal phase space (LPS) with coordinates $(\zeta, \delta)$ where $\zeta$ refers to the axial position of an electron with respect to the bunch's center and $\delta\equiv {p}/{\mean{p}}-1\simeq \Delta p_z/\mean{p_z}$ is the fractional momentum offset of an electron; here $\mean{p}$ represents the bunch mean momentum ($p_z$ refers to the longitudinal momentum). The axial field associated to the wakefield generated by the electron bunch is given by $E_z(\zeta)=\int_{-\infty}^{\zeta} \Lambda(\zeta-\zeta') \sum_{n,m} w^{(m)}_{n}\cos(k^{(m)}_n\zeta') d\zeta'$, where the double sum is evaluated on the number of modes $n=1,N$ categorized as monopole ($m=0$) and dipole ($m=1$) modes supported by the DLW. The parameters $w^{(m)}_n$ and $k^{(m)}_n$ are  respectively the field amplitude and wave vector associated to the mode $(n,m)$, and $\Lambda(\zeta)\equiv \int_{-\infty}^{+\infty} d\delta \Phi(\zeta,\delta)$ [here $\Phi(\zeta,\delta)$ is the LPS density distribution] is the charge density with the total bunch charge given by $Q=\int_{-\infty}^{+\infty} d\zeta \Lambda(\zeta)$. For sake of simplicity we only consider the dominant monopole ($m=0$) mode $n=1$ and introduce $w\equiv w^{(0)}_1$ and $k\equiv k^{(0)}_1$.  As an example we consider the case of a semi-Gaussian distribution  $\Lambda(\zeta)=\frac{Q}{\sqrt{2\pi\sigma^2}{\cal N}}[\exp(-(\zeta-\mu)^2/2\sigma^2)\Theta(\zeta-\mu) + \Theta(-\zeta+\mu)]$ where $\mu$ and $\sigma>0$ are respectively the rising edge center and rms width, ${\cal N}>0$ is a normalization constant, and $\Theta()$ the Heaviside function. Upon satisfying the transcendental condition, $\lambda =\frac{4\pi \sigma}{\sqrt{2}}D(\frac{\sqrt{2}\pi \sigma}{\lambda})$ (with solution $\sigma\simeq 5 \lambda$), where $D()$ is the Dawson function~\cite{dawson} and $\lambda \equiv 2\pi/k$ is the mode wavelength, the wakefield reduces to $E_z(\zeta)=\frac{Q}{{\cal N}} \exp\left(-\frac{2\pi^2 \sigma^2 N}{\lambda^2}\right)\cos[k (\zeta-\mu)]$,   Hence an initially smooth LPS distribution in $(\zeta_0,\delta_0)$ [Fig.~\ref{fig:setup}(b)] is energy modulated as it interacts with its wakefield over the length $l$ following $\delta_0 \rightarrow \delta_d= \delta_0(\zeta_0) + eV/\gamma_0 \cos(k\zeta_0+\psi)$ where $\gamma_0$ is the Lorentz factor, $\psi$ an arbitrary phase and $V$ the modulation potential (for the distribution above $V\equiv Q l /{\cal N}$ and $\psi\equiv -k \mu $); see Fig.~\ref{fig:setup}(c). The subsequent transport of the beam through the downstream drift space can be described by the linear transformation $\zeta_0 \rightarrow \zeta_f= \zeta_0 +\xi \delta_d$ where $\xi\simeq-L/\gamma_f^2$ is the longitudinal dispersion of a drift space with length $L$, and $\gamma_f$ is the Lorentz factor downstream of the DLW structure. A proper choice of $L$ and $\gamma_f$ leads to the energy modulation being converted into a density modulation at a given location downstream~\cite{lemery} and the density modulation period is equal to the mode wavelength $\Delta \zeta \simeq \lambda$; see Fig.~\ref{fig:setup}(d). 

The experiment was performed at the photoinjector test facility at DESY in Zeuthen (PITZ)~\cite{pitz} diagrammed in Fig.~\ref{fig:setup}(a). In brief, the $\sim 6.2$~MeV/c electron bunches are generated in an L-band (operating at a frequency $f=1.3$~GHz) radiofrequency (RF) photoemission electron source and directly focused into a DLW and further transported in a drift space up to an L-band linear accelerator (linac) where they are nominally accelerated to a final momentum of $\sim 20$~MeV/c~\cite{paramonov}. The RF gun comprises a high-quantum-efficiency Cesium Telluride photocathode illuminated by an ultraviolet laser pulse with a super-Gaussian temporal distribution. The temporal laser profile is produced via coherent pulse stacking using a \v{S}olc filter~\cite{shaper} with settings optimized to minimize density modulation on the electron beam. The DLW is located $z_c=1.71$-m from the photocathode and the solenoidal lenses surrounding the gun are tuned to focus the beam at the longitudinal center of the DLW.  The beam size at the center of the structure is measured to be $\sigma_{\perp}^*=102\pm 5$~$\mu$m for a bunch charge of  $Q=1.1\pm 0.05$~nC; the measurement was made by placing a Ce:YAG screen below the DLW holder on an actuator. Two DLW structures with different dimensions were available to our experiment. Both structures consist of a hollow fused-silica tube with its outer surface metalized with a copper layer of $\sim$1~$\mu$m (DLW1) or contacted to an aluminum support (DLW2). Downstream of the linac, a suite of beam diagnostics enables the measurement of the beam phase-space distribution and associated parameters. The accelerator settings relevant to the experiment along with the DLW parameters are summarized in Tab.~\ref{tab:settings}.
%
%
%
%
%
%
\begin{table}[t!!]
\caption{Settings of accelerator parameters relevant to the experiment. The listed value for the phases are offset with respect to the maximum momentum gain phase.\label{tab:settings} }
\begin{center}
\begin{tabular}{l c c c c}\hline\hline\
parameter & symbol & nominal & range & unit \\
\hline
laser launch phase & $\phi_l$  & 0 & -- & deg  \\
laser spot radius  & $r_l$  & 2 & -- & mm  \\
laser pulse duration & $L_t$ & 13 & $[10, 20]$ & ps \\
RF gun peak field  & $E_0$  & 60 & [45, 60] & MV/m  \\
linac phase& $\varphi_b$  & 0 & [-20, +10] & deg  \\
linac voltage & $V_b$  & 14 & [10, 18] & MV  \\
bunch charge  & $Q$ & 1.1 & [0.020, 2] & nC  \\
beam momentum & $\mean{p}$  & 21.8 & [16, 22] & MeV/c  \\
\hline 
DLW  permittivity & $\varepsilon_r$ & 4.41 & $-$ & $-$ \\
DLW1 inner radius & $a_1$ & $ 450 \pm 50 $  & $-$ & $\mu$m \\
DLW1 outer radius & $b_1$ & $ 550 \pm 50 $ & $-$ & $\mu$m \\
DLW1 length & $l_1$ & $50.0\pm 0.1 $ & $-$ & mm  \\
DLW2 inner radius & $a_2$ & $ 750 \pm 50 $  & $-$ & $\mu$m \\
DLW2 outer radius & $b_2$ & $ 900 \pm 50 $ & $-$ & $\mu$m \\
DLW2 length & $l_2$ & $80.0\pm 0.1 $ & $-$ & mm  \\
\hline
\hline
\end{tabular}
\end{center}
\end{table}

\begin{figure}[tt!!!!!!!!!!!!]
\centering
\includegraphics[width=0.46\textwidth]{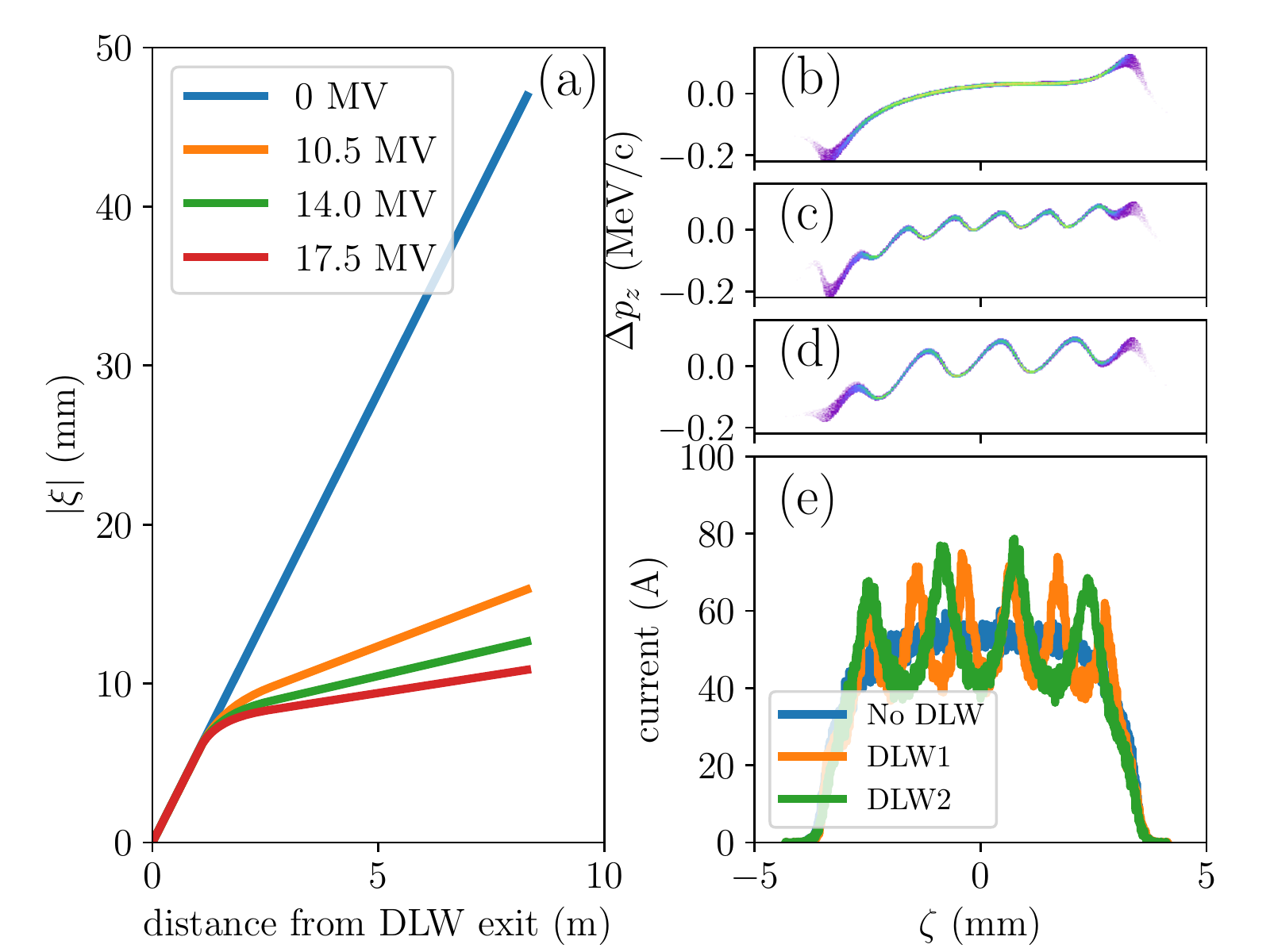}
\caption{Simulated evolution of the compression factor $|\xi|(z-z_c)$ along the beamline downstream of the DLW location for different booster-linac accelerating voltage $V_b$ (a) and final LPS distribution obtained for the nominal accelerator settings for the cases without DLW (b) and with DLW1 (c) or DLW2 (d) inserted along with corresponding current profiles (e).}
\label{fig:xi_and_Isim}
\end{figure}

In order to gain further insights on the experiment, we performed supporting numerical simulations of the beam dynamics using the program \astra~\cite{astra}. The software solves the equation of motion for electron macroparticles representing the bunch in the presence of externally-applied user-defined electromagnetic fields. The program also includes collective space-charge forces using a quasi-static mean-field treatment where the electrostatic field is computed in the bunch's reference frame using a particle-in-cell method; the Lorentz force in the laboratory frame is obtained from a superposition of the Lorentz-transformed space-charge and external fields. The electron-beam dynamics in the DLW is modeled using a Green's function approach detailed in~Ref. \cite{dohlus}. The Green's function is computed following the algorithm presented in Ref.~\cite{ng}; see Ref.~\cite{httpcode} for the associated software implementation. This model was used to produce the sequence of LPS snapshots displayed in Fig.~\ref{fig:setup}(b-d) and was benchmarked against a first-principle electrodynamics simulation performed with the software, {\sc echo}~\cite{echoz}. The corresponding beam parameters [root-mean square (rms) sizes,  emittances $\varepsilon_u \equiv 1/(mc) [\mean{u^2}\mean{p_u^2}-\mean{u p_u}^2]^{1/2}$ along the transverse ($u=\perp$) and longitudinal ($u=z$) degrees of freedom] are displayed in Fig.~\ref{fig:setup}(e-f) $-$ the beam is cylindrical symmetric. Additionally, the simulation allows for a numerical evaluation of the longitudinal dispersion taking into account the acceleration $\xi(z)=\int_{z_c}^{z} dz'/\gamma_f^2(z')$ downstream of the DLW (here $z>z_c=1.71$~m is the position along the beamline). It is especially found that $\xi$ increases slowly downstream of the linac thereby effectively ``freezing" the current profile. The latter effect is also supported by the evolution of the bunch current profile along the beamline; see Fig.~\ref{fig:setup}(g). For completeness the evolution of $\xi$ along the beamline appear in Fig.~\ref{fig:xi_and_Isim}(a) together with the final LPS and current distribution obtained for the various cases (no DLW, DLW1, and DLW2 inserted); see Fig.~\ref{fig:xi_and_Isim}(b-e). 
Additionally, the simulations were performed with and without considering the effect of the DLW and confirmed the minimal impact of the DLW on the transverse phase-space parameters as quantified by the negligible change on the transverse-emittance evolution; see Fig.~\ref{fig:setup}(e). It should be pointed out that the large transverse emittance excursions along the beamline are the results of a  non-optimum beamline configuration for the experiment. In practice, an optimized implementation of the passive bunching technique will  likely require additional focusing elements to control the beam size downstream of the DLW and mitigate the transverse emittance growth.  Finally, the longitudinal emittance is significantly increased at the DLW location due to the imparted energy modulation. It is however worth noting that the final longitudinal emittance downstream of the linac is only increased by $<5\%$ when the DLW is included compared to the case with no DLW; see Fig.~\ref{fig:setup}(e). 

The backbone diagnostics is an S-band ($f=2.997$~GHz) transverse deflecting structure (TDS) used to streak the beam~\cite{tds}. The TDS ($z=10.985$~m), vertically streaks the beam so that the vertical beam distribution measured on a Ce:YAG screen located $\sim 1.3$~m from the TDS centre is representative of the temporal bunch distribution; the vertical coordinate of an electron is related to its axial position via $y\simeq {\cal S} \zeta$ where the shearing parameter ${\cal S}$~\cite{klausTDS} is inferred from a beam-based calibration procedure. It should be noted that in the present experiment the temporal resolution of the streaking was limited to $\sim 0.5$~ps. A more precise value accounted for in the error analysis is obtained for each measurement via a beam-based procedure.

\begin{figure}[bb!!!!!!!!!!!!]
\centering
\includegraphics[width=0.46\textwidth]{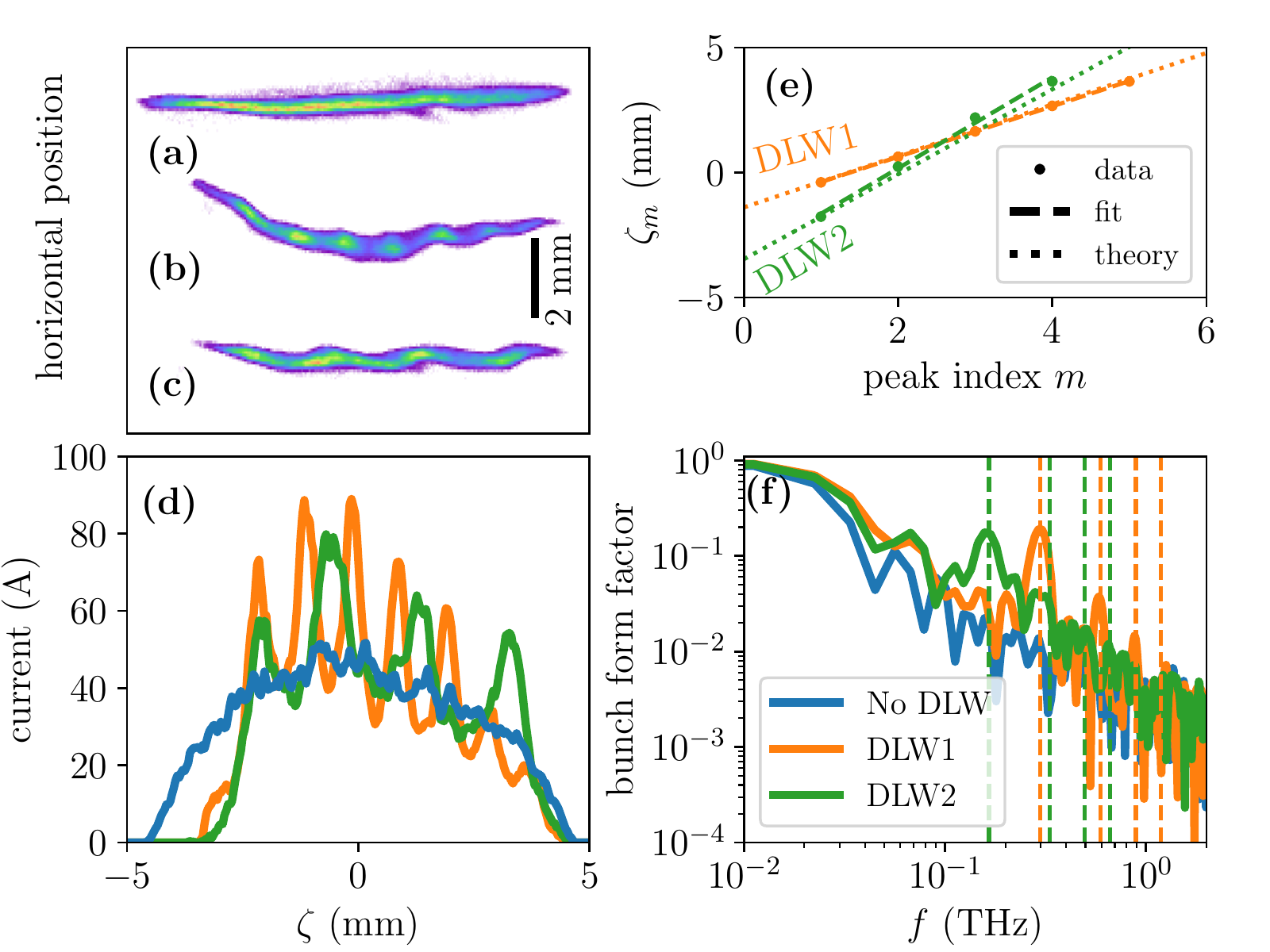}
\caption{Measured $Q(\zeta,x)$ charge-density distribution without DLW (a) and with DLW1 (b) or DLW2 (c) inserted, along with associated current profiles $I(\zeta)$ (d). Locations of local maxima for the current profiles measured with the DLW structures inserted (e) and associated bunch form factors (f).}
\label{fig:100A}
\end{figure}

The operating parameters of the RF gun and linac were further tuned to optimize the bunching process. Ultimately a $\sim 2$-fold peak-current enhancement was observed. The measured streaked density distributions appear in Fig.~\ref{fig:100A}(a-c) for the three cases under investigations (no DLW structure versus DLW1 or DLW2 structures inserted). The associated current profiles are displayed in Fig.~\ref{fig:100A}(d) and indicate that peak currents close to $\sim 90$~A are attained when the beam is propagated through a DLW. The observations are in qualitative agreement with the simulated current profiles; see Fig~\ref{fig:xi_and_Isim}(e): similar current-enhancement factors are measured  when the beam passes through one of the structures. The disagreement in absolute peak current is attributed to the lack of precise knowledge of the initial photocathode drive-laser temporal profile along with the possible contributions from other wakefield source which could change the overall correlation along the bunch and correspondingly affect the peak currents. 
To further quantify the origin of the observed modulation, the locations of the peaks  $\zeta_m=m\lambda_1+\zeta_{off}$ (where $m$ is an integer and $\zeta_{off}$ an arbitrary offset) are measured and a linear regression provides the wavelength of the modulation $\lambda_1$ which is dominated by the fundamental mode supported by the structure. The results of linear regressions give $\lambda_1^{DLW1}=1.01\pm 0.10$~mm and  $\lambda_1^{DLW2}=1.81\pm 0.10$~mm, in good agreement with the expected fundamental-mode wavelengths of $\lambda_1^{DLW1}=1.02\pm 0.16$~mm and  $\lambda_1^{DLW2}=1.58\pm 0.17$~mm respectively; see Fig.~\ref{fig:100A}(e). These values are obtained by directly solving the dispersion equation for the considered DLW with computed error bars accounting for the fabrication uncertainties listed in Tab.~\ref{tab:settings}. 

\begin{figure}[ttt!!!!!!!!!!!!]
\centering
\includegraphics[width=0.46\textwidth]{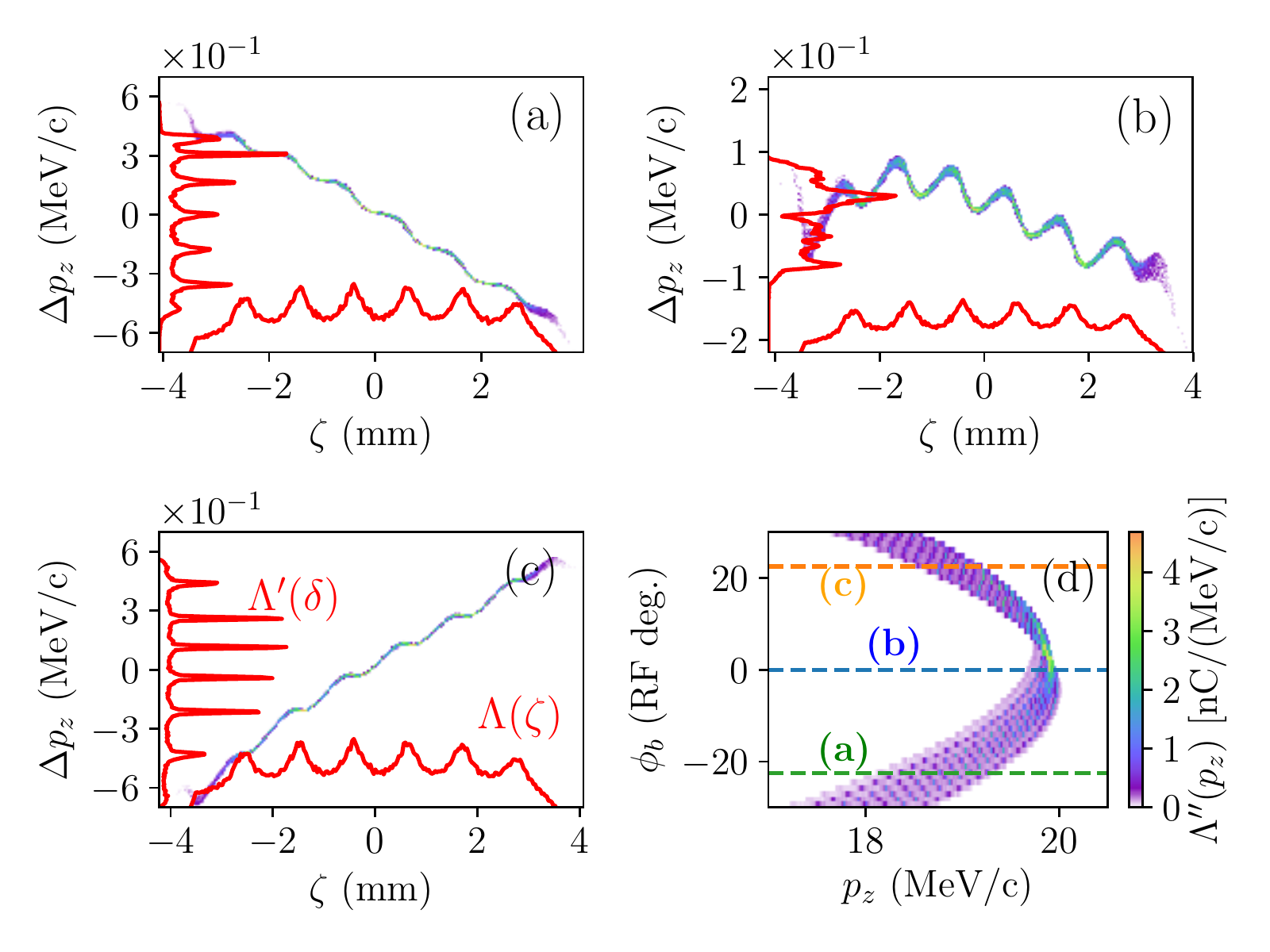}
\caption{Simulated evolution of the LPS for  booster linac injection phases $\varphi_b=-22.5^{\circ}$ (a), $0^{\circ}$ (b) and $+22.5^{\circ}$ (c). The traces correspond to the peak-normalized charge distribution as function of longitudinal (resp. energy) $\Lambda(\zeta)$  [resp. $\Lambda'(\delta)$] coordinate.  Evolution of the charge distribution $\Lambda''(p_z)$  for different injection phases $\varphi_b$ (d) with labeled lines referring to phase settings associated with plots (a), (b) and (c). The simulations are performed with DLW2 (similar results are obtained with DLW1 albeit with a different modulation period). 
\label{fig:lpsevolsimu}}          
\end{figure}

Finally, the individual peak durations can be further quantified by computing the bunch form factor (BFF) $b(f) \propto|\int_{-\infty}^{+\infty} dt I(\zeta/c) e^{-2\pi f\zeta/c}|^2$ of the current profile in the frequency ($f$) domain via a Fast Fourier Transform (FFT) algorithm; see Fig. ~\ref{fig:100A}(f). When one of the DLWs is inserted, the BFF displays the expected spectral enhancement at $f_1\simeq c/\lambda_1$ and at some of the harmonics frequencies $f_n=nf_1$ (where $n$ is an integer). DLW1 especially yields a spectral enhancement at the 3rd harmonic ($f_3\simeq 1$~THz) confirming the current modulations have a duration  $\tau < 1/f_3 \simeq 1$~ps, an upper value set by the resolution of the TDS-based technique. 
\begin{figure}[hhh!!!!!!!!!!!!]
\centering
\includegraphics[width=0.445\textwidth]{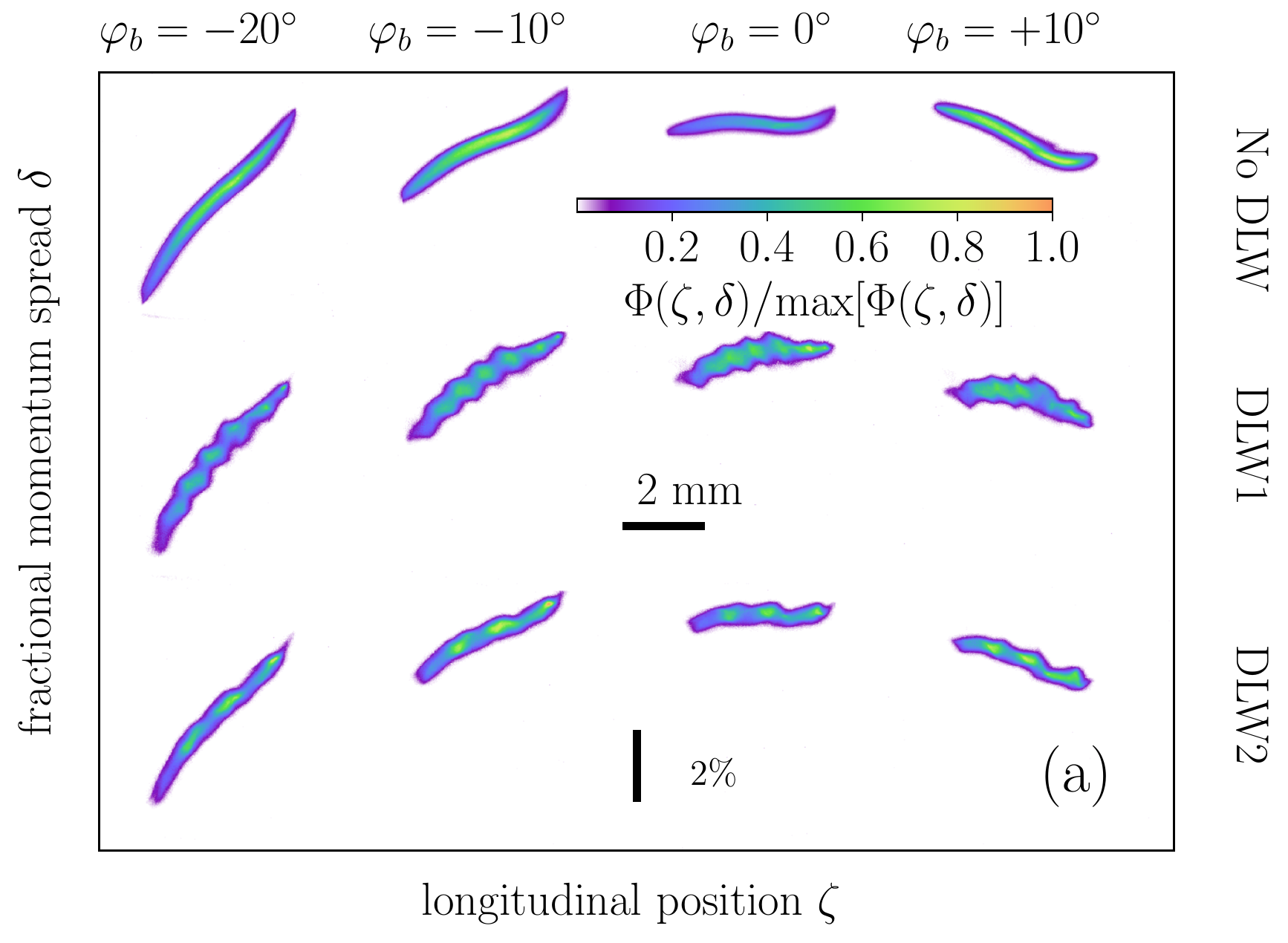}
\includegraphics[width=0.445\textwidth]{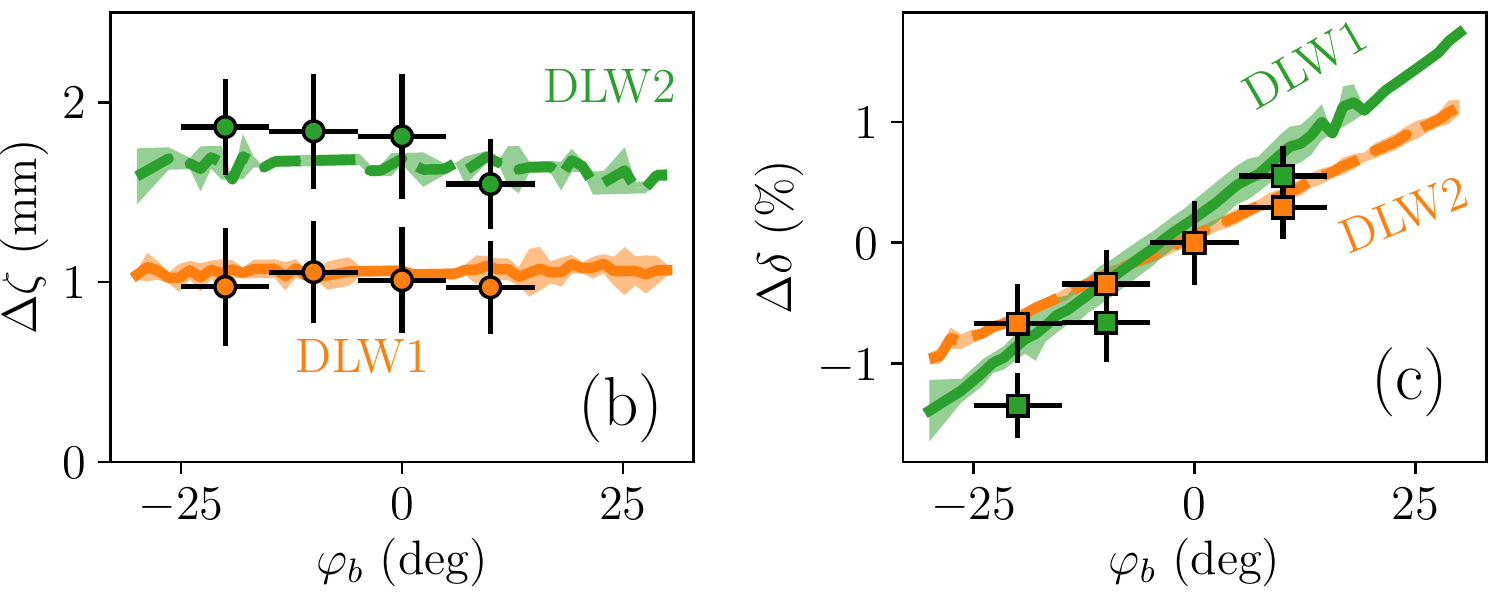}
\caption{Mosaic image of measured LPS-distribution snapshots (a) for different settings of the booster linac injection phases $\varphi_b$ (columns) corresponding to the three cases of DLW operation (rows). Comparison of the simulated (traces) and measured (symbols) modulation period along the longitudinal $\Delta\zeta$ (b) and fractional energy $\Delta\delta$ (c) coordinates. The shaded areas reflect uncertainties in the simulated values. 
\label{fig:lpsevol}}
\end{figure}

An important capability of the experimental configuration is the longitudinal-phase-space control enabled by the linac located downstream of the DLW, providing control on the final bunching configuration.  Operating the linac off-crest provides a knob to introduce a correlation between the time and energy coordinates. During acceleration through the booster linac the fractional momentum spread evolves as $\delta_f \rightarrow \delta_b=(1/\gamma_b)\{\gamma_f\delta_f + \Gamma_b [\cos(k_b \zeta_f +\varphi_b)-\cos(\varphi_b)]\}\simeq (\gamma_f/\gamma_b)\delta_f  +{\cal C} \zeta_f $ where the right-hand side approximation stems from the assumption $k_b \zeta \ll 1$, and $\gamma_b \equiv \gamma_f+\Gamma_b \cos(\varphi_b)$ is the final Lorentz factor downstream of the linac with $\Gamma_b\equiv eV_b/(mc^2)$ where $V_b$ is the booster-linac accelerating voltage. The booster wave-vector amplitude is $k_b=27.3$~m$^{-1}$.  Therefore off-crest ($\varphi_b \ne 0$) operation imposes a linear correlation  ${\cal C}\equiv - k_b\Gamma_b/\gamma_b  \sin\varphi_b$ within the LPS. 

The introduced LPS correlation can be taken advantage of to control, e.g., the energy of each microbunch within the beam as demonstrated via numerical simulations in Fig.~\ref{fig:lpsevolsimu}. Given that the bunch is accelerated, the longitudinal motion is unaffected by the phase of the booster and the temporal modulation is solely set by the DLW parameters. To demonstrate this LPS-control feature, we further propagated the vertically-streaked beam to a horizontally energy-dispersive beamline and measured the beam distribution on a downstream Ce:YAG screen ($z=20.885$~m). Under proper optimization, the coordinates of an electron are given by $y={\cal S}' \zeta$ (where ${\cal S}'\ne {\cal S}$) and $x=\eta \delta$ (where $\eta\simeq 0.9$~m is the dispersion function at the observation point) thereby enabling a direct measurement of the LPS density distribution. Figure ~\ref{fig:lpsevol}(a) displays snapshots of the LPS-density distribution measured for the three configurations and for four sets of the booster-cavity phase and qualitatively illustrates the control over the $\zeta-\delta$ correlation along the bunch.  The LPS-measurement is limited and we therefore measure the location of the LPS peaks value to infer the longitudinal $\Delta\zeta$ and energy $\Delta \delta$ separations between the peaks; see Fig.~\ref{fig:lpsevol} (b,c). The data is in agreement with the simulations and confirm that  tuning the phase $\varphi_b$ controls the energy separation between the microbunches while not affecting their longitudinal separation resulting in a tunable correlation between the microbunchs. Consequently, the control enabled by $\phi_b$ together with the ability to insert different structures provides a method to tailor the microbunch energy and longitudinal spacings. Such a versatile manipulation technique could have applications to multi-color free-electron lasers~\cite{roussel}  or ultra-fast electron diffraction.

In summary, we have demonstrated the basic features of a simple method to passively form a modulated beam by exploiting the beam-induced electromagnetic wakefields produced in a dielectric-lined waveguide; we note the concept could work with other high-impedance mediums also e.g. corrugated structures or plasmas. Although our observation leads to a modest peak current enhancement of a  factor $\sim2$, our simulations indicate the concept could in principle be applied to produce kA-class peak currents in an optimized beamline~\cite{lemery}. Additionally, for lower-energy beams and longer DLW structures, the bunching may occur within the structure and yield the emission of coherent Cherenkov radiation~\cite{cook} akin to a single-pass free-electron laser process~\cite{stupakov}. The simplicity and compactness of the demonstrated technique together with its versatility (it can be coupled to any electron-emission process) are appealing features that should motivate its implementation in compact electron sources being developed in support to fundamental research or various societal applications. Finally,  it should be pointed out that the method could be extended to compress a larger portion of the bunch (instead of introducing a density modulation) by selecting a DLW structure with a fundamental-mode wavelength comparable to the bunch length~\cite{lemery}. 

F.L. and P.P. express their gratitude to the PITZ team for their hospitality and excellent support. We thank M. Figora and A. Shultz (NIU) for manufacturing the DLW holder, and D. Zhang (CFEL, DESY) for help with the metallic coating of DLW1. This work was supported by the European Union's Horizon 2020 Research and Innovation programme under Grant Agreement No. 730871, by the European Research Council (ERC) (FP/2007-2013)/ERC Grant agreement No. 609920, by the US Department of Energy under contract No. DE-SC0011831 with NIU, by the German Bundesministerium f\"ur Bildung und Forschung, Land Berlin and by the Helmholtz Association.


\begin{thebibliography}{9}
\bibitem{englandPRL} R. J. England, J. B. Rosenzweig, and G. Travish, 
\href{https://doi.org/10.1103/PhysRevLett.100.214802}{Phys. Rev. Lett. {\bf 100}, 214802 (2008)}. 
\bibitem{muggliPRL} P. Muggli, et al. 
\href{https://doi.org/10.1103/PhysRevLett.101.054801}{Phys. Rev. Lett. {\bf 101}, 054801 (2008)}.
\bibitem{sunPRL} Y.-E Sun, et al, 
\href{https://doi.org/10.1103/PhysRevLett.105.234801}{Phys. Rev. Lett. {\bf 105}, 234801 (2010)}. 
\bibitem{piotPRL} P. Piot, et al.,  
\href{https://doi.org/10.1103/PhysRevLett.108.034801}{Phys. Rev. Lett. {\bf 108} 034801 (2012)}. 
\bibitem{haPRL} G. Ha, et al., 
\href{https://doi.org/10.1103/PhysRevLett.118.104801}{Phys. Rev. Lett. {\bf 118} 104801 (2017)}. 
\bibitem{antipov} S. Antipov, et al., 
\href{https://doi.org/10.1103/PhysRevLett.111.134802}{Phys. Rev. Lett. {\bf 111}, 134802 (2013)}. 
\bibitem{lemery} F. Lemery and P. Piot, 
\href{https://doi.org/10.1103/PhysRevSTAB.17.112804}{Phys. Rev. Accel. Beams {\bf 17}, 112804 (2014)}. 
\bibitem{andonian} G. Andonian, et al., 
\href{https://doi.org/10.1103/PhysRevLett.118.054802}{Phys. Rev. Lett. {\bf 118}, 054802 (2017)}. 
\bibitem{carlsten} B. E. Carlsten and S. M. Russel, 
\href{https://doi.org/10.1103/PhysRevE.53.R2072}{Phys. Rev. E {\bf 53}, R2072 (1996)}. 
\bibitem{BC} R. Talman,  
\href{https://doi.org/10.1103/PhysRevLett.56.1429}{Phys. Rev. Lett. {\bf 56}, 1429 (1986)}. 
\bibitem{piotFEL15} P. Piot,  
\href{http://accelconf.web.cern.ch/AccelConf/FEL2015/papers/mod02.pdf}{Proceedings of FEL15, Daejeon, Korea, p. 274 (2015)}. 
\bibitem{gover} A. Gover, 
\href{https://doi.org/10.1103/PhysRevSTAB.8.030701}{Phys. Rev. Accel. Beams {\bf 8}, 030701 (2005)}. 
\bibitem{XFELTHz} E. A. Schneidmiller, et al., 
\href{https://doi.org/10.1117/12.2017014}{Proc. SPIE 8778, Advances in X-ray Free-Electron Lasers II, 877811 (3 May 2013)}. 
\bibitem{ued0} A. H. Zewail and J. M. Thomas, {\em 4D Electron Microscopy: Imaging in Space and Time} (Imperial College Press, London, 2010).
\bibitem{ued} R. K. Li, et al., \href{https://doi.org/10.1063/1.3646465}{J. Appl. Phys. {\bf 110}, 074512 (2011)}. 
\bibitem{dawson} H. G. Dawson, \href{https://doi.org/10.1112/plms/s1-29.1.519}{Proc. Lond. Math. Soc. {\bf s1-29}, 519 (1897)}. 
%
\bibitem{pitz} M. Krasilnikov et al.,
\href{https://doi.org/10.1103/PhysRevSTAB.15.100701}{Phys. Rev. Accel. Beams {\bf 15}, 100701 (2012)}. 
\bibitem{paramonov} V. Paramonov et. al.,
\href{http://accelconf.web.cern.ch/AccelConf/LINAC2010/papers/mop081.pdf}{Proceedings of LINAC2010, Japan MOP81.}
\bibitem{shaper} I. Will and G. Klemz, 
\href{https://doi.org/10.1364/OE.16.014922}{Opt. Exp. {\bf 16} 14922 (2008)}. 
\bibitem{astra}  K. Fl\"ottmann, {\em {\sc astra}: A space charge algorithm, User's Manual}, (unpublished). 
\bibitem{dohlus} M. Dohlus, K. Fl\"ottmann, C. Henning,  
\href{https://arxiv.org/abs/1201.5270}{arXiv:1201.5270 [physics.acc-ph] (2012)}. 
\bibitem{ng} M.~Rosing, and W.~Gai, 
\href{https://doi.org/10.1103/PhysRevD.42.1829}{Phys. Rev. D {\bf {42}}, 1829 (1990)}. 
\bibitem{httpcode} P. Piot, F. Lemery, and J. Wang, Wakefield in Axi-Symmetric Dielectric-Lined Waveguides, {\sc python} program at \href{https://github.com/NIUaard/DiWakeCyl}{https://github.com/NIUaard/DiWakeCyl}
\bibitem{echoz} I. Zagorodnov,  
\href{https://doi.org/10.18429/JACoW-NAPAC2016-WEA1IO02}{Proceedings of NAPAC16, Chicago, IL USA, p. 642 (2016)}. 
\bibitem{tds} E. N. Volobuev, et al., 
\href{https://doi.org/10.1088/1742-6596/747/1/012080}{Journal of Physics: Conference Series {\bf 747}, 012083 (2016)}. 
\bibitem{klausTDS} K. Fl\"ottmann, and V. Paramonov, 
\href{https://doi.org/10.1103/PhysRevSTAB.17.024001}{Phys. Rev. Accel. Beams {\bf 17}, 024001 (2014)}. 
\bibitem{roussel} E. Roussel, et al., 
\href{https://doi.org/10.1103/PhysRevLett.115.214801}{Phys Rev Lett. {\bf 115}, 214801 (2015)}. 
\bibitem{cook} A. M. Cook, et al., 
\href{https://doi.org/10.1103/PhysRevLett.103.095003}{Phys. Rev. Lett. {\bf 103}, 095003 (2009)}. 
\bibitem{stupakov} G. Stupakov, 
\href{https://doi.org/10.1103/PhysRevSTAB.18.030709}{Phys. Rev. Accel. Beams {\bf 18}, 030709 (2015)}. 
\end{thebibliography}
\end{document}